# POLYPHONIC SOUND EVENT DETECTION FOR HIGHLY DENSE BIRDSONG SCENES

*Alberto García Arroba Parrilla[1]*

a.g.a.p.albertogarciaarrobaparrilla@tilburguniversity.edu

1. Tilburg University, the Netherlands

*Dan Stowell[1,2]*

d.stowell@tilburguniversity.edu

2. Naturalis Biodiversity Center, Leiden, the Netherlands

## ABSTRACT

One hour before sunrise, one can experience the dawn chorus where birds from different species sing together. In this scenario, high levels of polyphony, as in the number of overlapping sound sources, are prone to happen resulting in a complex acoustic outcome. Sound Event Detection (SED) tasks analyze acoustic scenarios in order to identify the occurring events and their respective temporal information. However, highly dense scenarios can be hard to process and have not been studied in depth. Here we show, using a Convolutional Recurrent Neural Network (CRNN), how birdsong polyphonic scenarios can be detected when dealing with higher polyphony and how effectively this type of model can face a very dense scene with up to 10 overlapping birds. We found that models trained with denser examples (i.e., higher polyphony) learn at a similar rate as models that used simpler samples in their training set. Additionally, the model trained with the densest samples maintained a consistent score for all polyphonies, while the model trained with the least dense samples degraded as the polyphony increased. Our results demonstrate that highly dense acoustic scenarios can be dealt with using CRNNs. We expect that this study serves as a starting point for working on highly populated bird scenarios such as dawn chorus or other dense acoustic problems.

**Index points** - Sound Event Detection, Polyphony, Birdsong, Dense scene

## 1. INTRODUCTION

Among other domains, acoustic biodiversity studies have flourished as improvements in data science have helped researchers understand, monitor and study many species in detail [1] [2] [3] [4]. Sound event detection tasks have covered many domains where relevant information can be extracted from acoustic scenarios [5] [6] [7]. SED problems can increase in difficulty when many classes are considered and when several events overlap at the same time [8] [9], also known as polyphony. SED models that can detect overlapping sound events can be addressed as polyphonic SED.

Polyphony is very common in real-life birdsong scenarios and requires polyphonic SED models to detect all overlapping events. To solve this problem, it requires a model capable of detecting, despite high levels of overlap, all existing species in the studied area. This addresses scenarios like dawn chorus, defined by [10] as a phenomenon around one hour before sunrise with a high activity of birdsong from many individuals and species. This could reach up to 30+ species [11] presenting a very dense scenario for studying and monitoring wildlife.

Our work aims to build a polyphonic SED model capable of dealing with birdsong real-life scenarios. We focus on two research questions (RQs):

- RQ1: What level of polyphony can be successfully detected on a dense audio scene and dense species list?
- RQ2: How accurately can this model detect very dense scenes with up to 10 overlapping sounds?

To address these research questions, several datasets with different polyphony were created and used to train SED models. This work demonstrates that a single neural network can provide decent results on low-polyphonic scenes and high-polyphonic ones, considering the complexity of the task.

## 2. RELATED WORK

On the one hand, a consensus on architecture and features is clear, as very good results have been obtained applying a commonly used design of CRNN [2] [5] [6] [9] [12] [13] [14] [15]. However, reviewing related papers, it can surely be stated that not much research has been carried out in analyzing high polyphony in the birdsong audio domain. Some papers analyze polyphony in other kinds of scenes, mostly urban sounds, [6] [9] [13] [16] or using non-specific datasets while others employ avian datasets but without giving much focus on dense polyphonic scenes [1] [3] [4] [17].

The most closely related project is the study of polyphony using a convolutional recurrent network of [14]. The author presents a model that can address birdsong detection with good results for recordings with up to 3 overlapping sounds. However, it has limitations as only a total of 5 species are considered making this approach not very close to a real-life dawn chorus scenario. Also related to birdsong, [18] proposed an HMM-based method to determine the temporal information of bird sounds without establishing any intrinsic limit on the polyphony. However, this study does not reflect how neural networks work on these problems.

In order to study dense polyphonic scenes, datasets should include a certain number of classes, but most of the datasets used in the literature do not get over 16. Many papers have employed data synthesis to force a certain level of polyphony in the samples but if the number of classes is limited, then the polyphony will be too.

Table 1: F score and error rate of the three models O3, O6 and O10 using their testing sets and 3 additional subsets with exactly 3, 6 and 10 overlapping events

|  | Matched testing set, with polyphony 1 to O_ | | Samples with polyphony of 3 | | Samples with polyphony of 6 | | Samples with polyphony of 10 | |
| --- | --- | --- | --- | --- | --- | --- | --- | --- |
|  | F Score | ER | F Score | ER | F Score | ER | F Score | ER |
| Model O3 | 0.49±0.17 | 0.61±0.19 | 0.45±0.18 | 0.65±0.21 | 0.32±0.14 | 0.77±0.11 | 0.30±0.13 | 0.79±0.12 |
| Model O6 | **0.51±0.15** | **0.60±0.15** | **0.52±0.18** | **0.57±0.20** | **0.47±0.13** | 0.64±0.12 | 0.41±0.09 | 0.72±0.08 |
| Model O10 | 0.47±0.11 | 0.65±0.18 | 0.46±0.14 | 0.88±0.40 | 0.46±0.12 | **0.62±0.11** | **0.45±0.09** | **0.68±0.07** |

Polyphony approaching a level of 30 or more [11] appears to be beyond the current state of the art, both in datasets and SED systems. Out work builds towards this goal, by training and evaluating an SED system for high-density polyphonic scenes typical in birdsong.

## 3. METHODS

### 3.1. Dataset election

The chosen dataset [19], contains 77 recordings of dawn chorus from the USA contemplating 48 species and a total of 16,052 event activations. Every audio file has a fixed sample rate of 32kHz, fixed length of 5 minutes and has an associated annotation file with very detailed information about every event.

A 48-species multi-class multi-label model presents a very complex SED problem, so after studying the distribution of the data where many classes have less than 10 annotations in the whole dataset, only those species with at least 100 activations were considered for the detection model, resulting in 20 species. This process reduced the complexity while still allowing to increase the polyphony up to 10.

### 3.2. Preprocessing and subsets creation

In order to feed the recordings to the model, frame blocking was carried out where the audio signal was split into analysis frames of 5-second-fixed length, shifted with a fixed hop size of 2.5 seconds. Additionally, data augmentation methods such as time stretch, pitch shift and shift were applied to help with generalization.

Three subsets were created for polyphony study of 3, 6 and 10 using data synthesis where several 5-second clips were selected and merged. Every created subset contains audio scenes where each clip has the same number of overlapping events (or less) as the determined polyphony. Annotations were also merged by simply concatenating files.

As an input feature, a NxM Mel spectrogram was calculated with a window size of 1024, a hop length half that size and 128 Mel bands where N is the number of frames and M the Mel bands. The annotation is transformed into a SxN binary (presence-absence) matrix where S denotes the number of classes.

### 3.3. Model

The model used is a Convolutional Recurrent Neural Network (CRNN) inspired by [9]. Firstly, the convolutional part consists of 5 layers following an evolution of increasing features in the order of: (64, 128, 128, 128, 264). After every convolutional step: layer normalization, leaky *ReLu* activation function and 2D max pooling on the frequency axis. Then, two bidirectional gated recurrent units (GRU) with 128 neurons each, with tanh activation and dropout of 0.5. And, finally, a time-distributed fully connected part of 2 dense layers of 128 neurons each using Leaky *ReLu* as activation function and a follow-up dropout of 0.5.

Adam was used as the model's optimizer with a fixed learning rate of 0.001 and binary cross-entropy was calculated as loss function.

Every different polyphonic model was trained with 10,000 training samples and 200 epochs, using a training batch size of 128 and the proportion of training-testing sets of 90-10. Three models were created and fitted using subsets of different polyphony as training. From this point, the models are referred to as O3, O6 and O10, whose numeric IDs correspond to the maximum polyphony found in their training subsets.

Every section of the model and processing of the data was carried out using *Python*, and libraries such as *Keras* and *sklearn*. The code for the preprocessing, model and evaluation is available in github.com/SrArroba/hd_sed .

### 3.4. Evaluation metrics

An evaluation metric system designed for SED tasks called *sed_eval* [20] is used to measure the performance of the models. F-score and error rate segment-based metrics were used to evaluate the models with a segment size ($t\_collar$) of 0.1 seconds. For every model, 4 subsets are used for testing: 3 subsets with fixed polyphony of 3, 6 and 10, respectively, and a testing set matching each model's training session conditions (i.e., files with up to the specified value of polyphony).

After performing a prediction, the output is binarized using 0.5 as thresholding value, as seen in Fig. 1, where the x-axis represents the time frames (up to 313 frames in 5 seconds) while the y-frame contains the 20 classes/bird species alphabetically sorted. Thresholding can be used to have a better understanding of the models' behavior and pick out further anomalies or faulty tendencies.

Table 2: Individual F score and ER for the 3 most common and 3 least common bird species in the dataset.

| Species | Number of annotations | % of annotations | Mean F Score | Mean ER |
|---|---|---|---|---|
| Eastern towhee (EATO) | 5238 | 36.63 % | 0.73±0.23 | 0.53±0.25 |
| Wood thrush (WOTH) | 1826 | 11.38 % | 0.42±0.38 | 0.51±0.36 |
| Northern cardinal (NOCA) | 884 | 5.51 % | 0.24±0.37 | 0.58±0.67 |
| Hermit thrush (HETH) | 162 | 1.01 % | 0 | 0.12±0.33 |
| Red-bellied woodpecker (RBWO) | 145 | 0.90 % | 0.07±0.08 | 0.07±0.25 |
| Blue-winged warbler (BAWW) | 142 | 0.88 % | 0 | 0.12±0.32 |

After this task, the predicted and ground truth can be compared to evaluate the model.

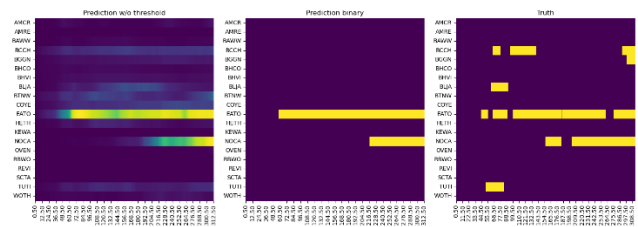

Fig. 1: Output after a sample is being run through the model. From left to right: prediction obtained with values between 0 and 1; the same prediction after applying a threshold of 0.5; ground truth.

## 4. RESULTS

While training the models, the loss was measured to study their evolution and how efficiently the models learnt. The first 100 epochs were recorded and shown together in Fig. 2.

Analyzing Table 1, it can be observed that O6 performs generally better than the other two models except on higher levels of polyphony than it was trained for, while the O10 model provides consistent results for all types of samples. Despite its complexity, the CRNN model performs adequately matching the conclusions reached by [15] about the fitness of CRNN over other networks to deal with event detection.

Additionally, F score and error rate were calculated for each bird species individually to check if there is a correlation between the model's performance on detecting them and their number of activations in the original dataset. Table 2 shows the number of annotations and evaluation metrics for the 3 most common and 3 least common species.

## 5. DISCUSSION

The goal of this study was to find a model that distinguishes bird species from a large list and temporally detects them when dealing with a dense scene of birdsong. Additionally, it aims to study whether a model can detect the species of a scene with up to 10 overlapping sounds (i.e., very dense scenario).

It could be hypothesized that models that trained with denser samples, learn faster than those who work with simpler scenarios. However, as seen in Figure 2, every model learns in a similar way while the range of their loss evolutions depends on how complex the samples are: the more complex, the higher the range of loss evolution due to the difficulty to learn denser scenes. Generally, models trained with denser scenes are very consistent for all densities while the model trained with simpler samples shows a degradation as the complexity increases. Additionally, models trained with denser scenes perform similar or better in simpler scenes. On the other hand, O10 gets worse results than the O6 model when dealing with emptier scenes. This could be because of the distribution of the training data regarding their polyphony.

Considering how every synthetic file is generated for each subset, it can denote that the higher the polyphony, the denser the scene and, therefore, more examples per class can be obtained in a single sample. This means that the O3 subset will have fewer representing features of the classes in a single input compared to O6 or O10. However, this does not mean that the models with denser samples learn faster as can be seen in Figure 8, where all models provide a similar evolution in their loss but at different values range. This difference can be explained as the more complex samples are, the more difficult to learn, and therefore, the higher the loss.

On the other hand, as shown in Table 1, models with denser scenes (such as O10) are very consistent for all densities while the model that learns with simpler samples, O3, shows a degradation as the complexity increases, having very poor performance on samples with 10 overlapping sounds. On the other hand, denser scenes mean more complexity, which slows the learning process of the model in that type of scenario. This can be observed in the O10 model where more complex samples are used for training but gets worse results than the O6 model when dealing with emptier scenes. Generally, models trained with denser scenes perform in a similar or better way on simpler samples.

While calculating the performance for each species, it could be hypothesized that it holds a correlation with the total number of activations found in the original dataset. Performing an evaluation using specific species it was demonstrated in Table 2 that species with an elevated number of activations provide better evaluation metrics when studied individually while species with very low appearance in the dataset gives significantly worse performance

Further work could be carried out investigating deeper architectures or more innovative feature selection to achieve better results on denser scenarios. When dealing with higher levels of polyphonies, CNN filters and GRU units should also increase, as suggested by [9].

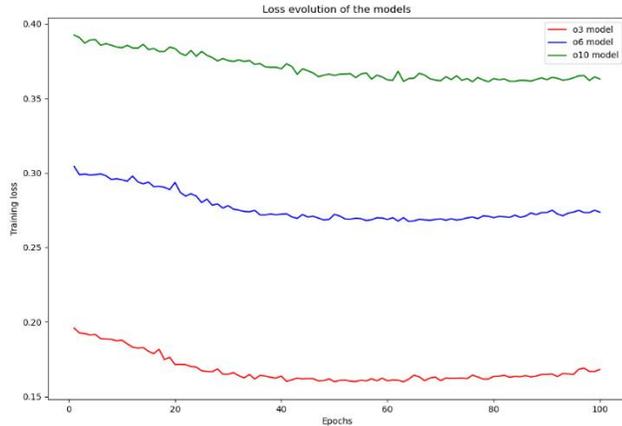

Fig. 2: Training loss for the first 100 epochs. The evolution is shown together for the three established models: O3, O6, O10.

## 6. CONCLUSION

In this project, a birdsong dataset of 20 classes was used to train several SED models capable of labeling and temporally allocating the species of birds present on highly dense scenes. Addressing the research questions initially stated, the three models created: O3, O6 and O10 performed suitably considering the complexity of the task.

It could be observed that every model displayed a similar learning curve, nevertheless, using too dense samples could affect the performance of the model when dealing with less crowded scenes but maintaining consistent results in all scenarios. Models' accuracy decayed when dealing with samples more complex than the ones used in training. Additionally, species that have more activations in the original dataset obtained the best results when studying them separately

This work serves to prove the feasibility of elements of modern design, in terms of architecture or features, when facing a task of detecting dense acoustic scenarios. Moreover, it also serves as motivation for other researchers to try new creative ways of dealing with this problem and as a challenge to complete it by adding a location task (SELD).

## 7. REFERENCES

[1] V. Lostanlen, J. Salamon, A. Farnsworth, S. Kelling & J. Bello, «Robust sound event detection in bioacoustic sensor networks,» PLoS ONE, pp. volume 14, issue 10, Pe0214168 2019 DOI: 10.1371/journal.pone.0214168, 2019.

[2] S. Madhusudhana, Y. Shiu, H. Klinck, E. Fleishman, X. Liu, E. Nosal, T. Helble, D. Cholewiak, D. Gillespie, A. Sirovi, et al., «Improve automatic detection of animal call sequences with temporal context,» Journal of the Royal Society Interface, 18(180):20210297, 2021.

[3] D. Stowell, M. D. Wood, H. Pamuła, Y. Stylianou & H. Glotin, «Automatic acoustic detection of birds through deep learning: The first bird audio detection challenge.,» Methods in Ecology and Evolution, vol. 10, no. 3, pp. 368–380, 2019.

[4] S. Kahl, T. Wilhelm-Stein, H. Klinch, D. Kowerko & M. Eibl, «Recognizing birds from sound - the 2018 birdclef baseline system,» CoRR, vol. arXiv: 1804.07177, 2018.

[5] E. Cakir, G. Parascandolo, T. Heittola, H. Huttunen & T. Virtanen, «Convolutional recurrent neural networks for polyphonic sound event detection,» IEEE/ACM Transactions on Audio, Speech, and Language Processing, 25(6), 1291-1303, 2017.

[6] Y. Cao, Q. Kong, T. Iqbal & F. An, «Polyphonic sound event detection and localization using a two-stage strategy,» arXiv:1905.00268, 2019.

[7] T. Heittola, «Computational Audio Content Analysis in Everyday Environments,» Tampere University Dissertations - Tampereen yliopiston väitöskirjat; Vol. 434, 2021.

[8] J. Abeßer, «USM-SED-A Dataset for Polyphonic Sound Event Detection in Urban Sound Monitoring Scenarios,» arXiv:2105.02592, 2021.

[9] S. Adavanne, A. Politis & T. Virtanen, «Multichannel sound event detection using 3D convolutional neural networks for learning inter-channel features,» international joint conference on neural networks (IJCNN) (pp. 1-7), 2018.

[10] P. R. Marler & H. Slabbekoorn, «Nature's music: the science of birdsong,» Elsevier, 2004.

[11] A. Farina, M. Ceraulo, C. Bobryk, N. Pieretti, E. Quinci & E. Lattanzi, «Spatial and temporal variation of bird dawn chorus and successive acoustic morning activity in a Mediterranean landscape,» Bioacoustics, 24(3), 269-288, 2015.

[12] S. Adavanne, A. Politis & T. Virtanen, «Localization, detection and tracking of multiple moving sound sources with a convolutional recurrent neural network,» arXiv:1904.12769, 2019.

[13] E. Cakir & T. Virtanen, «Convolutional recurrent neural networks for rare sound event detection,» Deep Neural Networks for Sound Event Detection, 12., 2019.

[14] M. Crous, «Polyphonic Bird Sound Event Detection With Convolutional Recurrent Neural Networks,» 10.13140/RG.2.2.11943.09126, 2019.

[15] R. Lu & Z. Duan, «Bidirectional GRU for sound event detection,» DCASE, 2017.

[16] K. Drossos, S. I. Mimilakis, S. Gharib, Y. Li & T. Virtanen, «Sound event detection with depthwise separable and dilated convolutions,» International Joint Conference on Neural Networks (IJCNN), 2020.

[17] T. Pellegrini, «Densely connected CNNs for bird audio detection,» 25th European Signal Processing Conference (EUSIPCO), 2017.

[18] D. Stowell & D. Clayton, «Acoustic event detection for multiple overlapping similar sources,» IEEE Workshop on Applications of Signal Processing to Audio and Acoustics (WASPAA), 2015.

[19] L. M. Chronister, T. A. Rhinehart, A. Place & J. Kitzes, «An annotated set of audio recordings of Eastern North American birds containing frequency, time, and species information,» 2021.

[20] A. Mesaros, T. Heittola & T. Virtanen, «Metrics for polyphonic sound event detection» Applied Sciences, vol. 6, no. 6, pp. 162–178, 2016.